\pdfoutput=1
\documentclass[preprint]{elsarticle} 

\usepackage[utf8]{inputenc}
\usepackage[T1]{fontenc}
\usepackage[english]{babel}

\usepackage{amsmath}
\usepackage{amsfonts}
\usepackage{multicol}
\usepackage{url}
\usepackage{dirtytalk}

\usepackage{booktabs}

\usepackage{xcolor}

\usepackage{tikz}
\usepackage{pgf}
\usepackage{pgfplotstable}
\usepackage{pgfplots}

\usetikzlibrary{arrows.meta}
\usepgfplotslibrary{groupplots}
\usepgfplotslibrary{dateplot}

\pgfplotsset{
	compat=newest,
	measurement/.style={
		width=110mm,
		height=40mm,
		grid style={dotted, black!50},
		table/col sep=comma,
		date coordinates in=x,
		scaled x ticks=false,
		max space between ticks=25pt,
		scale only axis=true,
		every axis plot/.append style={very thick},
		tick style={black, thick},
		grid=both,
		minor x tick num=3,
		axis x line = bottom,
	    axis y line = left,
	    xminorticks=true,
		enlarge x limits = 0.025,
	    xticklabel = {\year},
	    minor x tick num=3,
		xtick distance=.1,
	    date ZERO={1991-01-01},
	    xmin={1991-01-01},
		xmax={2021-01-01},
		minor xtick = {1991-01-01, 1992-01-01, 1993-01-01, 1994-01-01, 1995-01-01, 1996-01-01, 1997-01-01, 1998-01-01, 1999-01-01, 2000-01-01, 2001-01-01, 2002-01-01, 2003-01-1, 2004-01-01, 2005-01-01, 2006-01-01, 2007-01-01, 2008-01-01, 2009-01-01, 2010-01-01, 2011-01-01, 2012-01-01, 2013-01-01, 2014-01-01, 2015-01-01, 2016-01-01, 2017-01-01, 2018-01-01, 2019-01-01, 2020-01-01},
		xtick={1991-01-01, 1994-01-01, 1997-01-01, 2000-01-01, 2003-01-01, 2006-01-01, 2009-01-01, 2012-01-01, 2015-01-01, 2018-01-01, 2021-01-01},
	}
}

\definecolor[named]{ACMBlue}{cmyk}{1,0.1,0,0.1}
\definecolor[named]{ACMYellow}{cmyk}{0,0.16,1,0}
\definecolor[named]{ACMOrange}{cmyk}{0,0.42,1,0.01}
\definecolor[named]{ACMRed}{cmyk}{0,0.90,0.86,0}
\definecolor[named]{ACMLightBlue}{cmyk}{0.49,0.01,0,0}
\definecolor[named]{ACMGreen}{cmyk}{0.20,0,1,0.19}
\definecolor[named]{ACMPurple}{cmyk}{0.55,1,0,0.15}
\definecolor[named]{ACMDarkBlue}{cmyk}{1,0.58,0,0.21}

\newcommand{\FigureLink}{%
    \tikz[x=1.2ex, y=1.2ex, baseline=-0.05ex]{%
        \begin{scope}[x=1ex, y=1ex]
            \clip (-0.1,-0.1) 
                --++ (-0, 1.2) 
                --++ (0.6, 0) 
                --++ (0, -0.6) 
                --++ (0.6, 0) 
                --++ (0, -1);
            \path[draw, 
                line width = 0.5, 
                rounded corners=0.5] 
                (0,0) rectangle (1,1);
        \end{scope}
        \path[draw, line width = 0.5] (0.5, 0.5) 
            -- (1, 1);
        \path[draw, line width = 0.5] (0.6, 1) 
            -- (1, 1) -- (1, 0.6);
        }
    }

\usepackage{hyperref}
\usepackage{multirow}

\usepackage[framemethod=TikZ]{mdframed}

\usepackage{xpatch}
\makeatletter
\xpatchcmd{\endmdframed}
  {\aftergroup\endmdf@trivlist\color@endgroup}
  {\endmdf@trivlist\color@endgroup\@doendpe}
  {}{}
\makeatother

\newcommand{\figureref}[1]{\begin{flushright}\FigureLink~Figure~\ref{#1}\end{flushright}}

\newcounter{observation}[section]
\renewcommand{\theobservation}{\arabic{observation}}

\newenvironment{observation}[1]{
	\refstepcounter{observation}
	\begin{mdframed}[
		frametitle={\colorbox{white}{\space Observation \theobservation\space}},
		innertopmargin=0pt,
		frametitleaboveskip=-\ht\strutbox,
		frametitlealignment=\raggedright,
		nobreak=true,
		skipabove=10pt,
		skipbelow=10pt,
	]%
	\label{#1}}{\end{mdframed}}

\newcounter{hypothesis}[section]
\renewcommand{\thehypothesis}{\arabic{hypothesis}}

\newenvironment{hypothesis}[1]{
	\refstepcounter{hypothesis}
	\begin{mdframed}[
		frametitle={\colorbox{white}{\space Hypothesis \thehypothesis\space}},
		innertopmargin=0pt,
		frametitleaboveskip=-\ht\strutbox,
		frametitlealignment=\raggedright,
		nobreak=true,
		skipabove=10pt,
		skipbelow=10pt,
	]%
	\label{#1}}{\end{mdframed}}

\newcommand{\orcidlink}[1]{}

\journal{Information and Software Technology}

\begin{document}

\begin{frontmatter}

\title{A Replication Study on Measuring the Growth of Open Source} 

\author[bth]{Michael Dorner\texorpdfstring{\corref{cor}\orcidlink{0000-0001-8879-6450}}}
\ead{michael.dorner@bth.se} 
\address[bth]{Blekinge Tekniska Högskola, Valhallavägen 1, 37179 Karlskrona, Sweden}
\cortext[cor]{Corresponding author}

\author[fau]{Maximilian Capraro\orcidlink{0000-0002-7598-6615}}
\ead{maximilian.capraro@fau.de} 
\address[fau]{Friedrich-Alexander-University Erlangen-Nürnberg, Computer Science, Schloßplatz 4, 91054 Erlangen, Germany}

\author[uoc]{Ann Barcomb\orcidlink{0000-0003-2126-9511}}
\ead{ann@barcomb.org} 
\address[uoc]{University of Calgary, Schulich School of Engineering, 2500 University Drive N.W., T2N 1N4 Calgary, Canada}

\author[bth]{Krzysztof Wnuk\orcidlink{0000-0003-3567-9300}}
\ead{krzysztof.wnuk@bth.se}

\begin{abstract}
\noindent\textbf{Context:} Over the last decades, open-source software has pervaded the software industry and has become one of the key pillars in software engineering. The incomparable growth of open source reflected that pervasion: Prior work described open source as a whole to be growing linearly, polynomially, or even exponentially.

\noindent\textbf{Objective:} In this study, we explore the long-term growth of open source and corroborating previous findings by replicating previous studies on measuring the growth of open source projects.  

\noindent\textbf{Method:} We replicate four existing measurements on the growth of open source on a sample of 172,833 open-source projects using Open Hub as the measurement system: We analyzed lines of code, commits, new projects, and the number of open-source contributors over the last 30 years in the known open-source universe. 

\noindent\textbf{Results:} We found growth of open source to be exhausted: After an initial exponential growth, all measurements show a monotonic downwards trend since its peak in 2013. None of the existing growth models could stand the test of time. 

\noindent\textbf{Conclusion:} Our results raise more questions on the growth of open source and the representativeness of Open Hub as a proxy for describing open source. We discuss multiple interpretations for our observations and encourage further research using alternative data sets.
\end{abstract}


\newpageafter{abstract}

\end{frontmatter}

\section{Introduction}

Open-source software is pervasive in modern software development: Open-source development tools help to build software, and open-source components are used as part of other software. Open source has evolved from small communities of volunteers driven by non-monetary incentives to foundations that host large projects and support decentralized innovation among many global industries \cite{Fitzgerald2006}. 

Most studies on open-source projects are vertical and have investigated single, often extraordinary, successful, or special open-source projects \cite{Godfrey2004, Robles2005, Roy2006, Succi2001}, or, on occasion, software forges \cite{Squire2017}. Three horizontal, longitudinal studies investigated the growth of open source as a whole: \cite{Capiluppi2003} from 2003, \cite{Koch2007} from 2007, and in 2008 \cite{Deshpande2008}. All three studies are sample studies aiming for generalizability over the population of open-source projects which means the focus is not on specific contextual details \cite{Stol2018}. From those three prior studies, open source is expected to grow \cite{Capiluppi2003}, grow quadratically \cite{Koch2007}, or grow exponentially \cite{Deshpande2008} for lines of code and number of projects. But obviously, no system with a polynomial or even exponential growth rate can remain stable: At some point, it might plateau or reach saturation because all available resources or needs are exhausted. 

Replications serve an important purpose in empirical software engineering as they broaden our understanding of what results hold under what conditions \cite{Shull2008}. We designed and executed an \emph{exact} replication in which the procedures of an experiment are followed as closely as possible \cite{Shull2008} from the three previous horizontal studies about the growth of open source and applied them to a new and updated set of open-source projects. Our goal is to understand long-term trends and implications \cite{Brooks2008, Shull2008}. Our main goal is to enable meta-analysis \cite{Kitchenham2020} of these and our experiment to gain more knowledge about open-source software. 


In detail, the contributions of this paper are: 

\begin{itemize}
	\item a detailed discussion of the three prior studies,
	\item the multi-dimensional measurements using Open Hub as a measuring system to quantify the growth of open source with respect to lines of code, commits, contributors, and projects, and, thereby, 
	\item the dependent and independent replication of the measurements by three prior studies.
\end{itemize}

Throughout this paper, we follow the terminology of metrology as defined in \cite{isovim} and the terminology on dependent and independent replications as defined in \cite{Shull2008}.

The following structure guides the reader through the study: In Section~\ref{sec_state_of_the_art}, we analyze all three studies and extract their measurement results as hypotheses. In Section~\ref{sec_measurement_method}, we describe our measurement method, including the used measurands, Open Hub as measuring system, and our filtering approach ensuring the correctness of the data. In Sections~\ref{sec_results} and \ref{sec_limitations}, we present and discuss the results of our measurements, while in Section~\ref{sec_discussion} we consider possible interpretations of our findings and discuss future work. In the last section, we summarize and conclude our findings.

\section{State of the art}
\label{sec_state_of_the_art}

In this study, we replicate the measurements of three studies: \cite{Capiluppi2003} from 2003 (Study~A), \cite{Koch2007} from 2007 (Study~B), and \cite{Deshpande2008} from 2008 (Study~C). Table \ref{tab_replicated_studies} lists all three studies with their time of publishing, their source, the number of open-source projects in the sample, and the considered time frames. 

\begin{table*} \centering 
	\begin{tabular}{|l|l|l|l|l|l|l|}
		\hline
		{Study} & {Year} & {Source} & {Projects} & {Time frame} \\ \hline
		A \cite{Capiluppi2003} & 2003 & FreshMeat.net & 406 & unknown -- 2002-07-01 \\ 
		B \cite{Koch2007} & 2007 & SourceForge.net & 4,047 & unknown \\ 
		C \cite{Deshpande2008} & 2008 & Ohloh.net & 5,122 & 1995-01-01 -- 2006-12-31 \\ \hline
	\end{tabular}
	
	\caption{Three prior \cite{Capiluppi2003,Deshpande2008,Koch2007} studies on the evolution of open source, showing data source, project sample size, and the considered data time frame.}
	\label{tab_replicated_studies}
\end{table*}

From those three prior works, we extracted four measurements and hypotheses, which are either explicit or implicit hypotheses, which are listed in Table~\ref{tab_hypotheses}.

\begin{table*} \centering
	\begin{tabular}{|ll|l|l|}
	\hline
	\multicolumn{2}{|l|}{Hypothesis} & {Study} \\ \hline
	{1} & Open source grows with respect to byte size. & A \\
	{2} & Open source grows quadratically with respect to lines of code. & B \\
	{3} & Open source grows exponentially with respect to lines of code. & C \\
	{4} & Open source grows exponentially with respect to projects. & C \\ \hline
	\end{tabular}%
	
	\caption{All hypotheses extracted from studies A, B, and C, along with the types of measurements from this study.}
	\label{tab_hypotheses}
\end{table*}

The original data sets of all studies were not publicly archived by the authors and, in the case of Studies~A and B, the original data sources no longer exist.

\subsection{Study~A}

Study~A \cite{Capiluppi2003} from 2003 presented a descriptive longitudinal study on 406 open-source projects and analyzed 12 different quantities of open-source projects. Although measured three times (February 2001, January 1st, 2002, and July 1st, 2002), all quantities are statically analyzed containing the whole time frame only and are not presented continuously over time or at least over those three or four sampling points. The exact sampling point (1999) and the date of first data collection (also 1999 or February 2001) remains unclear.

A pseudo-sampling was applied: the study uses the project status defined by the FreshMeat platform, an index for open-source software, with the status \emph{planning}, \emph{pre-alpha}, \emph{alpha}, \emph{beta}, \emph{stable}, \emph{mature}. It is unclear how this status is determined. For each possible status, half of the projects were randomly selected, resulting in 406 projects, according to the authors half of the population in 1999 hosted on FreshMeat. However, the sample could be drawn from ``living'' or inactive open-source projects, or both. Neither living nor inactive is defined. 

The original data source FreshMeat does not exist any more: On October 29, 2011, FreshMeat was renamed to Freecode. Since June 18, 2014, Freecode is no longer maintained. The quantities age, application domain, programming language, size, number of developers, number of users, modularity level, documentation level, popularity, and vitality rely on computations by the portal owner or are derived from those. Their construction is unclear and, in consequence, we cannot replicate the measurements.

Additionally, the measurements for vitality and popularity are explicitly discussed in the study. According to this study, 23\% of the projects increased their vitality. Vitality $V$ is defined by

\begin{equation}\label{eq_vitality}
	V = \frac{R \cdot A}{L}
\end{equation}

where---according to description---$R$ is the number of releases in a given period ($t$), $A$ is the age of the project in days, and $L$ is the number of releases in the period $t$. We assume a typo in the formula and/or the definitions: From the definition, $R = L$ and, therefore, $V = A$ in equation \ref{eq_vitality}. We also miss the portion of constant vitality to estimate the growth. 

90\% of projects did not change their status within six months. How the status is computed/defined and what are the six months remains unclear. 

60 projects out of 400 (15\%) are active; the rest are considered lethargic. Study~A defines an active project to have an increasing \say{vitality, popularity and subscribers and developers}. All projects which are not active are defined as lethargic.  It remains unclear if all three measurands must be increasing ($\land$) or at least one of them ($\lor$) for a project to be deemed active.

Popularity $P$ is defined by

\begin{equation}\label{eq_popularity}
	P = \sqrt[3]{\frac{U \cdot R \cdot (S+1)}{3}}
\end{equation}

where $U$ stands for the count of visits to the project homepage, $R$ is the number of visits to the project on FreshMeat pages, and $S$ is the number of subscribers. It is unclear to what extent the study controlled for bots and web crawlers, and what measures were taken to merge duplicate user identities. According to the description, $P$ is normalized between 0\% and 100\%, which is not true for any $U, R, S \geq 1$, unless an unspecified adjustment was subsequently applied.

Study~A measured the byte size of the source code excluding \say{documentation and unessential binary or code files, such as HTML, GIF, JPG}.  The study found that 63\% of the examined projects have not changed their size over six months, with 34\% of projects changing less than 1\%. In conclusion, only 3\% of all projects changed more than 1\% in size over six months. Over the longer timeframe encompassed by the first and last samplings, 59\% of projects did not change in size, 22\% grew by up to 10\%, 15\% grew between 10\%-50\% in size, and 5\% more thn doubled in size.  The authors observed that open source grows: 

\begin{hypothesis}{hypothesis_os_grows}
	Open source grows with respect to size (in bytes).
\end{hypothesis}

The authors neither discussed the magnitude of growth nor made predictions about future growth.

\subsection{Study~B}

Study~B \cite{Koch2007} from 2007 found the growth of open source to be best described as a quadratic function for a sample of 7,734,082 commits with 663,801,121 lines of code added and 87,405,383 removed from 8,621 projects contributed by 12,395 developers. The study did not discuss whether the lines of code count includes comments or not. Only CVS projects are considered. Neither Study~B nor the supplemental work \cite{Koch2004} contains any information on the time when the sampling took place and the considered time frame.

In detail, the authors found the growth better described as a quadratic 

\[
	S_B(t) = a \cdot t^2 + t \cdot b + c
\]

than as a linear function

\[
	S_A(t) = a \cdot t + b
\]

where $S_B$ and $S_A$, respectively, is the size in lines of code at time $t$ as days after the first commit. Both models are evaluated by the adjusted $R^2$ value, which is not applicable for non-linear regressions like this \cite{Spiess2010}. Study~B observed:

\begin{hypothesis}{hypothesis_os_grows_quadratically}
	Open source grows quadratically with respect to lines of code.
\end{hypothesis}

\subsection{Study~C}

Study~C \cite{Deshpande2008} from 2008 found an exponential growth of open source with respect to source lines of code, and number of new and total number of open-source projects. Comment lines and empty lines are excluded from the lines of code count. 

Study~C considers the 5,122 most popular open-source projects according to the number of in-links provided by the Yahoo! search engine to their website. A list of those open-source projects is not available to us. In contrast to Study~A and B, the original data source, Ohloh (now Open Hub) is still available. 

Study~C excluded \say{all commits where lines of code added is greater than average code added per commit plus three times the standard deviation}. Although mathematically not equivalent, we assume that the $z$-score as outlier detection is imitated. The exact measurement is not completely described in the paper: Although labeled in the plots, the description of the approach indicates that not lines added, but the net change with respect to lines of code is considered. The measurement of the total number of projects is redundant to the measurement of newly added projects because for a population growing at an exponential rate, the removal rate from the population must be smaller. This is a sufficient and necessary condition for exponential growth. However, if one considers lines of code added only, the lines of code added before the sampling period are removed and older projects are disadvantaged.

Study~C used $R^2$ to evaluate the goodness of the models, which is not an adequate measure for the goodness of fit in non-linear models \cite{Spiess2010}. Additionally, the curve fitting and its initial parameters are not presented. 

We conclude the following two hypotheses, which are derived from the findings of Study~C:

\begin{hypothesis}{hypothesis_os_grows_exponentially_loc}
	Open source grows exponentially with respect to source lines of code.
\end{hypothesis}

\begin{hypothesis}{hypothesis_os_grows_exponentially_projects}
	Open source grows exponentially with respect to the number of projects.
\end{hypothesis}

\section{Measurement method} \label{sec_measurement_method}

Since growth is a relative phenomenon, we focus also on relative measures: In this section, we describe the relative increase of lines of code, commits, and projects added to and persons contributing to the known open-source universe. 

Our measurements replicate the measurements from the prior studies, and, thereby, evaluate the extracted four hypotheses (Table \ref{tab_hypotheses}). Dependent replications are exact replications where researchers attempt to keep all the conditions of the experiment the same or very similar. Independent replications are also exact replications, where researchers deliberately vary one or more major aspects of the conditions of the experiment \cite{Shull2008}. Thus, we perform: 
	\begin{itemize}
		\item an independent replication of Study~A's measurement on size (in bytes) by measuring size in lines of code,
		\item a dependent replication of Study~B's measurement on LoC,
		\item a dependent replication of Study~C's measurement on LoC and projects,
		\item an independent replication of all previously mentioned measurements by measuring contributors, as we are measuring human activities
	\end{itemize}
	
In the following subsections, we define and discuss the four measurands, introduce Open Hub as our measuring system, and describe our filtering approach in detail.

\subsection{Measurands}

\subsubsection{Lines of Code}\label{loc}

A \emph{line of code} (LoC) is a non-blank line containing either comments, source code, or both. A \emph{source line of code} (SLoC) is a non-blank line starting with source code. A \emph{comment line of code} (CLoC) is a non-blank line containing comments only.\footnote{This definition also applies to inline comments such as \texttt{/* some comment*/ int i = 0;} in C/C$++$, also at the beginning.} 

In our study, we use added lines of code, a relative measurement. This derivative of LoC has its origin in \emph{diff}, a Unix command-line tool to calculate and display the line-based difference between two files \footnote{\url{https://github.com/blackducksoftware/ohcount/blob/master/src/diff.c\#L366}}. Open Hub evaluates the type of lines of code using \emph{ohcount}, which is publicly available\footnote{\url{https://github.com/blackducksoftware/ohcount}}. This tool can only distinguish between code and comment (and blank) lines. However, a line with both source code and followed by comment is classified as lines of source code---although having both. 

Study~B and C estimated the size of available source code in open source by measuring lines of code. Study~C excluded comment lines of code. It remains unclear if Study~B included or excluded comment lines. However, a comment is a valuable contribution to a software project. In modern programming languages like Go or Python, comments are directly embedded in the source code for documentation purposes. Therefore, we dependently replicate the measurements of both Study~B and C by measuring lines of code, which contains both comment and source lines of code. However, we also distinguish between source and comment lines of code during our measurements. 

Study~A measured the byte size of the source code excluding \say{documentation and unessential code, such as HTML, GIF, JPG}. Because the exact inclusion and exclusion criteria and definition of unessential code are not available or described by Study~A, we are independently replicating this measurement by measuring lines of code: Any non-empty line of code has at least one byte. Therefore, our measurement is consistent with Hypothesis~\ref{hypothesis_os_grows}, which states that open source grows (in bytes). By measuring lines of code, we can independently replicate the measurements of study A.

\subsubsection{Commits} \label{sec_commits}

Although lines of code is a popular and simple measurement, the choice of programming language, tooling, coding style guidelines, and embedded documentation has a large impact on lines of code. 
We believe that measuring commits are more suitable and robust for estimating the growth of open source in total, in particular with respect the growth of effort spent in open source.

A \emph{commit} is a semantically enclosed and author-tailored code contribution to a software project. Commit is the predominant term across most version control systems such as git, CSV, SVN, Mercurial, and Bazaar. Synonyms are, for example, \emph{change set} in Microsoft's Team Foundation Server (TFS). Unlike commits in data management (e.g., databases), commits in version control systems are persistent and kept in the repository. Each commit contains, in addition to other information, meta-information on 

\begin{itemize}
	\item the code change to be contributed,
	\item a unique identifier,
	\item the timestamp of the commit action, and
	\item the author of the commit.
\end{itemize}

Commit size calculated in lines of code has also been described as following a power-law distribution \cite{Arafat2009} or a Pareto distribution \cite{Hattori2008, Kolassa2013a}. The majority of commits affect 2-4 files, 6-46 lines of code, or 2-8 sections of contiguous lines of code \cite{Alali2008}.

Although the characteristics of a commit can vary depending on its context, tooling, and personal or project-specific preferences, several studies used commits for effort estimation \cite{Capiluppi2013, Tsunoda2006} or to measure collaboration among organizational boundaries \cite{Capraro2018}. Using commits as the basis for estimating effort is supported by research showing that the interval between consecutive commits does not vary widely \cite{Ma2014, Kolassa2013}, implying that developers typically put about the same amount of time into each commit. When commit size is calculated as the total number of commits in a period, it follows a power-law distribution \cite{Lin2013}. 

Therefore, we use the measurement of commits for independently replicating measurements of lines of code.

\subsubsection{Contributors}\label{sec_contributors}

Because all of the previous measurements are human-based activities (writing code, starting new open-source projects, or contributing a commit), we would also like to extend the existing measurements by measuring contributors and replicating all other measurements.

A \emph{contributor} is a role of a person who contributes to a project. In this paper, we focus exclusively on code contributions and accepted changes to the project's source code. In open source usually, only a committer can accept changes because only they have write-access to the repository. A human being can serve as multiple contributors, for example, depending on their professional affiliation. Open Hub can map contributor identities to a person. However, this mapping requires a registered user and is, in general, not trivial.

\subsubsection{Projects}\label{sec_project_states}

An \emph{open-source project} is a software project that complies the criteria defined by the Open Source Initiative{\footnote{\url{https://opensource.org/osd}}. The measurement on new open-source projects does not correlate with their identification our measurement system Open Hub: While the detection and identification by Open Hub can happen any time in their existence, the timestamp of their first commit remains the same and indicates the birth for an open-source project. 

In contrast to classical software projects, open-source projects do not have a pre-defined scope, thus an explicit beginning and an explicit end. Although an open-source project may not be actively developed anymore and abandoned, the code and all related artifacts such as documentation and communication may still be publicly available. Similarly, LoC added to a project tend to persist as long as they are not deleted or refactored, usually, projects remain open source. However, projects can be deleted deliberately or deleted when hosting platforms like Google Code are shut down and the project members do not actively migrate the project to a new platform. In ongoing research, we address the changes of the open-source projects within their lifecycles.

This measurement on new projects aims to replicate the measurement of projects by Study~C.

\subsection{Measuring system}

As our primary measurement system, we rely on Open Hub\footnote{We are not referring to the data collection tool with the same name described in \cite{Farah2014}.}, which was well-known under the name Ohloh until August 2014.
%
%
Open Hub is an online platform that provides an infrastructure to crawl and index open-source projects from different sources. Open Hub's crawlers support all main open-source version control systems: git, CSV, SVN, Bazaar, and Mercurial. This allows it to collect commit information across different source code hosts like GitHub, Gitlab, BitBucket, SourceForge, etc. The data source also includes deleted open-source projects and their commit history, which allows a retrospective view on the development of open source. 
 
In addition to Study~C, \cite{Nagappan2013} used Ohloh to develop a classification to assess the quality of a sample of open-source projects and to identify open-source projects that could be added to improve the sample quality. Their sample consisted of 20,028 active open-source projects, where active is defined as at least one commit per year. 

Open Hub provides all required information for this replication study: It stores monthly measurements on lines of code added to, commits contributed to, and contributors worked on the open-source projects in the observed open-source universe persistently. From the timestamp of the first commit, we can derive the birth of the open-source project. 

We crawled those data points through a REST API with XML responses from Open Hub for each project in the sample. Additionally, we gathered the project's Open Hub page as HTML to parse all duplicate warnings because this information is not available through the REST API. Our crawling and analysis toolchain as well as the anonymized data set is fully published on GitHub\footnote{\url{https://github.com/michaeldorner/quo-vadis-open-source}}.

\subsection{Filtering}\label{sec_filtering}

At the time of the data collection (2021-06-04 to 2021-06-07), Open Hub lists 355,111 open-source projects as of 2021-06-06. The development activity is available for 173,305 projects. This is the initial dataset. We limit our analysis to the last 30 years, exclude duplicated projects, and remove outliers in measurements. The following subsections describe those three steps of our filtering approach.

\subsubsection{Timeframe}

We limited the analysis time frame 1991-01-01 to 2020-12-31, inclusively. Incorrect time configuration on developers' machines can cause a wrong timestamp for the commit, either accidentally (e.g., Unix time, which is the Thursday, 1 January 1970) or on purpose. For example, according to its Git-history, the Go programming language started back in 1972 with a commit\footnote{commit hash \newline \texttt{7d7c6a97f815e9279d08cfaea7d5efb5e90695a8}} by Brian Kernighan with a C-code snippet -- obviously a remembrance for Kernighan's famous \textit{hello world} memo. Also, our measurement system Open Hub assigns 1970-01-01\footnote{\url{https://github.com/blackducksoftware/ohloh_scm/blob/ca0e512177fb9958473812445d6a54b551b3ce9b/lib/ohloh_scm/git/activity.rb\#L215}} if there is no valid timestamp. The default Unix timestamp (1970-01-01) is already excluded from the sample; we bypass this issue by limiting our analysis from the time frame 1991-01-01 to 2020-12-31. Those exceptions aside, we assume the timestamps to be correct by default. Smaller deviations (e.g., by setting up the time manually, wrong time zones, etc.) are smoothed by Open Hub's monthly sampling. 

After filtering our data set for the time frame from 1991-01-01 to 2020-12-31, 173,265 projects were available for further analysis.

\subsubsection{Duplicates}

Some projects contain commits that originate from other projects. We extracted the information on duplications from all the collected projects' HTML pages as it is not accessible through the REST API and excluded the projects marked as duplicates. We relied on the duplication detection by Open Hub, which is not publicly available.

We excluded 621 duplicates from 435 original projects. The Linux Kernel was the most duplicated project with 72 duplicates. Removing the duplicated projects results in 172,833 unique open-source projects in the data set.

\subsubsection{Outlier detection}

In our data set, we found outliers. For example, we found values up to 453,380 commits by one contributor in November 2018 within the \emph{Beagle Board} project\footnote{\url{https://github.com/beagleboard/beagleboard-org}}: A large-scale change to a database and to log files was split into tiny commits, each with some dozen added lines on one specific file. Those outliers are inherent in every kind of measurement and must be addressed by an outlier detection.

However, this study neither covers every open-source project nor aims to measure the absolute size of open source but its growth, a relative measure. Thus, excluding outliers does not limit the generalizability of our findings. Therefore, we detect and exclude error-prone measurements in this study by applying a statistically well-established $z$-score to all measurements of lines of code, commits, and contributors. The measurements on the new open-source projects are based on the initial sampling of the projects and the number of commits and are, therefore, not filtered separately. 

The $z$-score is defined by

\[z = \frac{x - \mu}{\sigma}\]

where $\mu$ is the mean of the population and $\sigma$ is the standard deviation of the population. $99.7 \%$ of all $z$-scores are greater than $-3$ and smaller than $+3$. All other values are commonly considered as outliers. 

We excluded for all four measurements---lines of code, commits, number of stateful projects, and contributors---all values with $|z| > 3$. Table \ref{tab_outliers} shows the outliers detected and their impact. 

\begin{table*} \centering
	\begin{tabular}{|l|l|l|l|} \hline
		{Measurement} & {Total} & {Sum of all outliers} & {Outlier threshold}\\ \hline
		Lines of code & 34,487,694,594 & 13,045,874,360 & 553,097\\
		Commits & 114,794,355 & 29,272,733 & 708\\
		Contributors & 8,549,545 & 2,542,617 & 26\\ \hline		
	\end{tabular}
	\caption{The outliers detected for each measurement within the considered timeframe without duplicates and its threshold value per month per project.}
	\label{tab_outliers}
\end{table*}

\section{Results}\label{sec_results}

In this section, we present the results of the analysis on the measurements of lines of code, commits, number of published projects, and contributors over 172,833 open-source projects in the sample. The measurement method is described in detail in section \ref{sec_measurement_method} and all collected data as well as code for measurement, preprocessing, and visualization is publicly available\footnote{\url{https://github.com/michaeldorner/quo-vadis-open-source}}.

For describing time frames, we define \emph{until} and \emph{since} to be inclusive. For example, until 2020 means that the timeframe ended on 2020-12-31 23:59:59.999. In all four measurements, we found an initial, transient exponential growth. This exponential growth is evaluated by using an exponential function $y = a \cdot \exp(b\cdot x) + c$ with the initial parameter $a, b, c = 0$ with a resolution of full years. Those exponential models are depicted in all relevant graphs by a dashed red line. The exact parameters are published in the related Jupyter notebooks. 

Before we discuss the findings of the individual measurement results in the following subsection, we would like to provide an overview of the three measurements on lines of code, commits, and contributors. Although none of the proposed growth patterns (linear, quadratic, or exponential) applies to our measurements,  lines of code, commits, and contributors follow the same growth pattern: After an initial exponential growth until 2010 and a peak in 2013, those measurements show a downwards trend. Figure \ref{fig_overview} provides a normalized comparison of the measurements lines of code, commits, and contributors. 

\begin{figure*}[!htp] \centering
	\resizebox{\textwidth}{!}{
	}
	\caption{The normalized measurements of lines of code, commits, and contributors follow the same growth pattern.}
	\label{fig_overview}
\end{figure*}

\begin{observation}{obs_same_pattern}
	The normalized measurement of lines of code, commits, and contributors follow the same growth pattern which cannot be described as one of the existing growth models. \figureref{fig_overview}
\end{observation}

%


\subsection{Lines of code}

As described in Section~\ref{loc} in detail, we consider in this study source and comment lines of code added. Some of those lines may be deleted or overwritten over time. Figure~\ref{fig_loc} depicts an aggregated view on lines of source code and comment lines of code over time, with the exponential model for lines of code. 

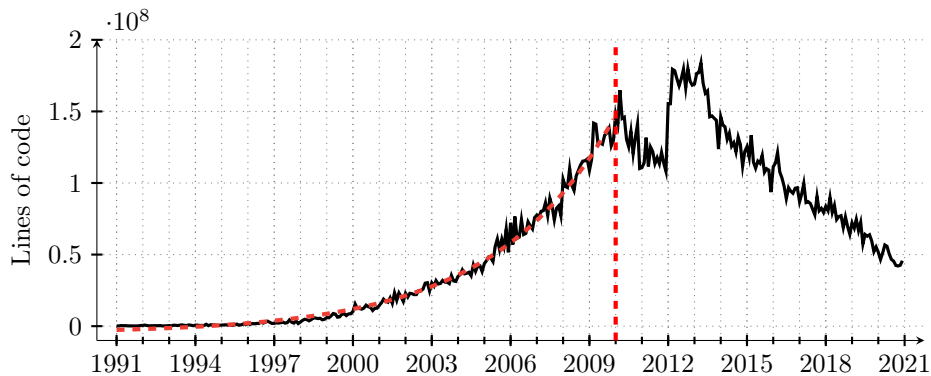
\begin{figure*}[!htp] \centering
	\begin{tikzpicture}
		\begin{axis}[measurement, ymin=-10000000, ymax=200000000, ylabel={Lines of code},]
		 
		\addplot[black] table[x={month}, y={loc_added}] {data.csv};
		\addplot[ACMRed, dashed, ultra thick] table[x={month}, y={exp_loc}, col sep=comma] {data.csv};
		
		\draw [red, dashed, ultra thick] ({2010-01-01}, -10000000) -- ({2010-01-01}, 200000000);
		
		\end{axis}
	\end{tikzpicture}
	\caption{Lines of code over time, showing lines of code and combined lines of comments and code. An exponential curve no longer fits the combined comments and code after 2013, while the source code growth rate declines after 2011.}
	\label{fig_loc}
\end{figure*}

\begin{observation}{obs_loc_decreases}
	Although initially growing exponentially until 2009, the growth in lines of code has continuously slowed down since 2013. \figureref{fig_loc}
\end{observation}

The number of lines of code and comments added to open-source projects slowed down and reached the level of the year 2005 at the end of 2020 after a global maximum in April 2013 with 183,831,021 source and comment lines added.

%

\subsection{Commits}

Again, we found an exponential growth of commits until 2010 and can confirm the measurement results of lines of code by replication. Figure~\ref{fig_commits} shows the total number of commits over time. 

\begin{figure*}[!htp] \centering
	\begin{tikzpicture}
		\begin{axis}[measurement, ymin=-50000, ymax=700000, ylabel={Commits},]
		 
		\addplot[black] table[x={month}, y={commits}] {data.csv};
		\addplot[ACMRed, dashed, ultra thick] table[x={month}, y={exp_commits}] {data.csv};
		
		\draw [red, dashed, ultra thick] ({2010-01-01}, -50000) -- ({2010-01-01}, 700000);
		
		\end{axis}
	\end{tikzpicture}

	\caption{New commits made over time. The exponential growth curve does not fit after the end of 2009.}
	\label{fig_commits}
\end{figure*}
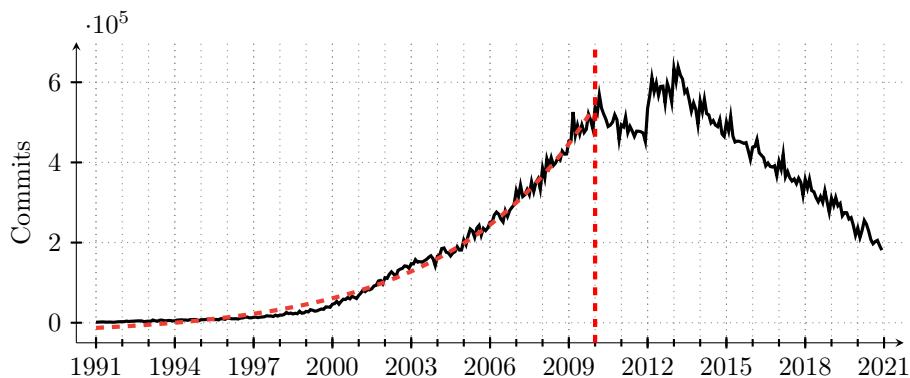

\begin{observation}{obs_commits_decreases}
	Although the number of open-source commits grew exponentially until 2009 and peaked with over 600,000 commits in March 2013, it has since declined. \figureref{fig_commits}
\end{observation}

After a dent in 2010 and 2013, the commits monthly contributed to open source reached its peak with more than 600,000 commits in March 2013. With seasonal minima in December from 2013 to 2019, the trend of commits contributed to open source is downwards. In 2020, it reached a level comparable to the end of 2005.

\subsection{Contributors}

Again, we observed exponential growth of open-source contributors until 2009. Figure~\ref{fig_contributors_over_time} illustrates our observations. In contrast to the other measurements, the number of contributors is not relative (new, added contributors), but absolute measurement. This measurement does not distinguish between different contributor types and their contribution frequency \cite{Barcomb2018}. 

\begin{figure*}[!htp] \centering
	\begin{tikzpicture}
		\begin{axis}[measurement, ymin=-5000, ymax=50000, ylabel={Contributors},]
		 
		\addplot[black] table[x={month}, y={contributors}] {data.csv};
		\addplot[ACMRed, dashed, ultra thick] table[x={month}, y={exp_contributors}, col sep=comma] {data.csv};
		
		\draw [red, dashed, ultra thick] ({2010-01-01}, -5000) -- ({2010-01-01}, 50000);
		
		\end{axis}
	\end{tikzpicture}
	\caption{Contributors per month over time. The exponential curve does not fit after end of 2009. The number of contributors peaked in 2013, and since 2014 the general trend has been a decline.}
	\label{fig_contributors_over_time}
\end{figure*}
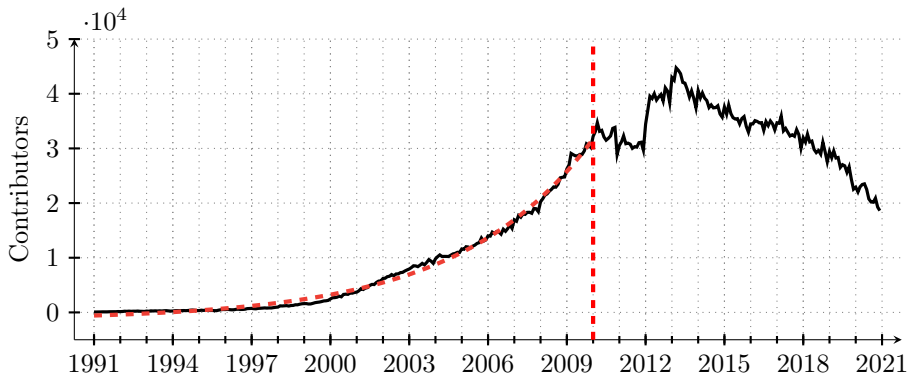

As for the monthly commits, the number of monthly contributors decreased after a peak in 2013. We also can see a yearly, seasonal drop in December around Christmas. 

\begin{observation}{obs_contributors_decrease}
	Although growing exponentially until 2009 and reaching its peak in March 2013 with 107,915 contributors, the number of open-source contributors has, as of 2018, decreased to the level of 2008. \figureref{fig_contributors_over_time}
\end{observation}

We refer the reader to Figure~\ref{fig_contributors_over_time} and the sharp rise in 2012. 

\begin{observation}{obs_erratic_contributor_rise}
	We observe an erratic rise of contributors in 2005, and again in 2012. \begin{flushright}\FigureLink Figure \ref{fig_contributors_over_time}\end{flushright}
\end{observation}

We cannot with certainty attribute the two sharp rises in contributors to any definite cause. However, it is possible that the factor mentioned as a source of measurement error in counting contributors (Section \ref{sec_contributors}), namely an increase in bot activity, is responsible. In a sample of GitHub projects, \cite{Wessel2018} identified a significant increase in bot adoption, beginning at a slightly later date, after 2013. Our sample includes but is not limited to GitHub projects, which might explain the difference. Other possible factors are the increase of paid developers in this period \cite{Riehle2014}, potentially with multiple professional affiliations, and the widespread shift from centralized to distributed version control (git was published in 2005), which changed how commits were attributed \cite{Rodriguez-Bustos2012}.

As elaborated in Section \ref{sec_contributors} on systematic errors of measuring contributors, the exact numbers must be treated with caution. Especially tracking individual contributors, and matching identities to persons are error-prone and became increasingly difficult with distributed version control systems such as git. However, we assume that the number of identities per user remains stable over time: Most open-source developers want to be identifiable and rewarded for their contributions. Both trends, the number of contributors, and the effort spent in open source (measured in commits per contributor) are downwards. The reduction in effort per person is consistent with research describing the growing recognition of episodic participation in open-source projects \cite{Barcomb2020}.

\subsection{Projects}




Also for the newly added projects to the observed open-source universe, we again are able to confirm an initial exponential growth until 2010, which does not continue. Figure~\ref{fig_projects} depicts the projects added to the observed open-source universe over time. 

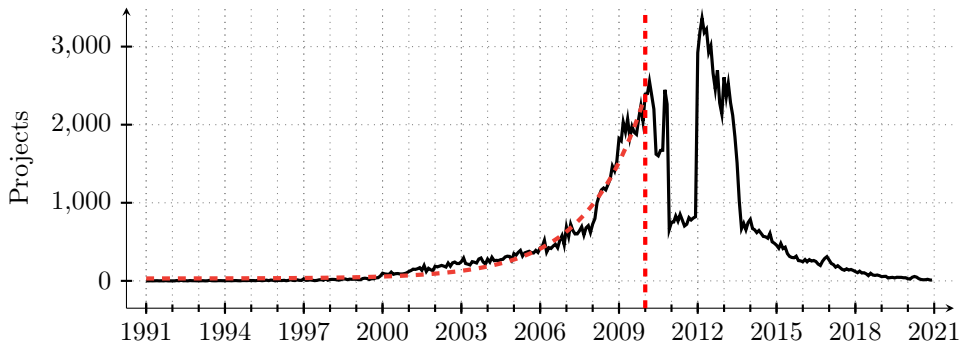
\begin{figure*}[!htp] \centering
	\begin{tikzpicture}
		\begin{axis}[measurement, ymin=-350, ymax=3500, ylabel={Projects},]
		 
		\addplot[black] table[x={month}, y={projects}] {data.csv};
		\addplot[ACMRed, dashed, ultra thick] table[x={month}, y={exp_new_projects}, col sep=comma] {data.csv};
		
		\draw [red, dashed, ultra thick] ({2010-01-01}, -350) -- ({2010-01-01}, 3500);
		
		\end{axis}
	\end{tikzpicture}
	\caption{New open-source projects per month over time. While the exponential curve fits until 2010, none of the existing growth models applies afterwards. }
	\label{fig_projects}
\end{figure*}

\begin{observation}{obs_projects_over_time}
	While the exponential curve fits until 2010, none of the existing growth models applies afterwards. \figureref{fig_projects}
\end{observation}

There is a significant drop in new projects in 2011, which we are not able to explain. As we use the timestamp of the first commit as indicator and not the project's identification by our measurement system, we assume no sampling issue but speculate that Open Hub has a tendency to under-represent newer, smaller projects although we were not able to prove our speculation. 

\begin{observation}{obs_new_projects_drop}
	We observe a significant decrease of new projects in the year 2011 and a sustained decline in 2013. \figureref{fig_projects}
\end{observation}

\section{Limitations}\label{sec_limitations} 

The most serious threat to the validity of our study is the unknown precision and accuracy of Open Hub as a measurement system. This could result both in a sample of open-source projects that is not representative of the universe of all open-source projects and has insufficient precision and accuracy of the measurements. In a first attempt, we evaluated the precision and accuracy measurements by Open Hub by manually measuring lines of code and commits on five projects based on maximum variation sample: very large projects (Linux Kernel, OpenStack), midsized projects (KDE, Golang), and small projects (BeeTee). We found no significant deviations beyond crawling jitter. However, this cannot be seen as a full evaluation as it is not statistically sound. A faulty measurement system Open Hub would also invalidate the findings of \cite{Nagappan2013} and, of course, Study~C \cite{Deshpande2008}. 

\citeauthor{Bruntink2014} assessed the quality of Ohloh in its version of the year 2013~\cite{Bruntink2014}. They observed improper (1) SVN configurations, (2) missing, and (3) inconsistent values. We also encountered the issue of projects which cannot be crawled due to outdated or wrong repository configuration (which we excluded). We also encountered similar problems while crawling Open Hub: 181,846 of 355,111 projects do not contain information on the development activity. We also found that the accumulated number of lines added does not fit the measured lines of code. Additionally, we found a large drop in added projects in 2011 we are not able to explain (Figure \ref{fig_projects}). We speculate that Open Hub could have decreased the number of projects it adds, so that newer projects are under-represented. 

Open Hub marks reported duplicates of open-source projects (mostly Linux kernel). How Open Hub detects duplications remains unclear. We assume further, undetected duplications. By excluding duplicates of the largest open-source projects (e.g., Linux kernel), we believe that we excluded a significant, though not a complete set of duplicates. In open source, forking is a common phenomenon. A fork is a secession from an existing open-source project to a new, independent development. This split applies not only to the software development itself but also to the developer community. The data set includes forks such as LibreOffice (forked from OpenOffice) but does not contain temporary forks such as those from GitHub. 
Although we excluded duplicated open-source projects from our analysis, we relied on the duplication detection by Open Hub, which is not publicly available.

Obviously, the choice of data source can affect findings. A recent study evaluating the Software Heritage Archive found public software growth rates to be exponential over more than 40 years \cite{rousseau2019growth}. The difference in findings can be potentially be explained in two ways that do not challenge the integrity of either data source. First, Open Hub includes only collaborative projects under an open-source license, while The Software Heritage Archive preserves all public code, including, for instance, example code and personal websites hosted on GitHub, in addition to non-code materials, such as collaborative writing projects hosted on GitHub. We assume that Software Heritage is a superset of open source, but the stake of open source is unknown. Second, Open Hub explicitly seeks to exclude duplicates, while the aforementioned study found that files and commits in multiple contexts---in other words, duplication---was a significant factor in the increase of source code. 
The same is true for GitHub, as of today one of the largest code hosting platforms and the related GHTorrent project \cite{Gousi13} which seems to be no longer maintained since June 2019. An unknown number of projects on GitHub is not intended to be an open-source projects aligning with the definition of open source and have no open-source license. 

We have good reasons to believe that the Open Hub measurement system is appropriate and applicable:
\begin{itemize}
	\item Other studies have also relied on Open Hub and its predecessor Ohloh \cite{Nagappan2013} assuming representativeness of Open Hub for open source---including one of the studies replicated \cite{Deshpande2008}. 
	\item Open Hub's observable and measurable open source universe contains 172,833 open-source projects (in January 2020) and is thus the largest collection of exclusively open-source projects we are aware of. 
	\item Different platforms (such as GitHub, Gitlab, BitBucket, etc.) and tools are covered.
	\item We assume the data set tends to be more towards large, vivid, and larger open-source projects, and to neglect code dumps, non-code open-source projects, and intentionally hidden open-source projects.
\end{itemize}

All measurements are based on the assumption that a human being contributed to open source. However, a contributor does not need to be a human being and a commit or line of code is not necessarily written by such. For example, at Google most of the commits come from bots \cite{Potvin2016}. The extent of bot-generated code in open source is unknown and beyond the scope of this study. Bots do not affect the measurement of created projects since bots do not kick off new projects, but all other measurements are potentially affected.

The number of rejected or abandoned contributions is unknown and not considered in this work. However, because open source became a mass phenomenon, the acceptance rate may not be constant, but decreasing. 

In our study, we refer to unavailable projects when we mean not reachable for Open Hub (anymore). This could also be the case for moved or archived projects or using a new version control system and not updating the Open Hub profile.  

Changing the version control system (e.g. Linux kernel started in 1991, but switched to \emph{git} in April 2005) can disturb the results. A change of version control system from centralized to distributed can also affect the perceived number of contributors because, in earlier systems, commits were often attributed to the committer, regardless of who authored the commit \cite{Rodriguez-Bustos2012}.

Our outlier detection and removal (see Section~\ref{sec_filtering}) neglect large open-source projects with more than 26 developers in a given month, more than 708 monthly commits, or more than 553,097 lines of code added per month: All large open-source projects like OpenStack, Linux, and Kubernetes are identified as outliers. Although we excluded important and stellar open-source projects, we decided to apply an outlier detection as discussed to have a more robust and statistically sound representation of open source. 



\section{Discussion}\label{sec_discussion}

We are surprised by our findings, although indications of stagnation in open-source contributions have also been described in qualitative studies \cite{Blind2021, Nagle2020}.

There are two possible reasons for our results: One explanation is a limitation of the data source. The other option is that the data are correct, and open source has reached a (temporary) plateau. We discuss each of these possibilities in more depth below.

If Open Hub is not a representative data source, this could be due to inherent flaws, which we have discussed in Section \ref{sec_limitations} along with the steps we have taken to ameliorate this concern. Another possibility is that open source is shifting from larger, centralized projects of the type tracked by Open Hub to smaller, more distributed projects. This would mean that growth in open source simply is not captured in a curated data source such as Open Hub. 

In the case of the findings being an accurate reflection of the state of open source, future research is needed to explain this change, as companies and open-source projects will need to adopt strategies that address the limited resources. While the data we investigated does not provide an answer, we can think of several potential, non-mutually exclusive explanations: 

\begin{itemize}
	\item A decrease in developers willing to volunteer, and no corresponding increase in paid development work
	\item The shift from volunteer to paid contributions reducing the effective time for contributing for each participant due to company resource management
	\item An increase in episodic participation \cite{Barcomb2018}, with more people preferring to volunteer less
	\item A generational shift (the mean age of contributors in 2005 was 31, and in 2017 it was 30 \cite{Ghosh2005, OS2017}) from collective (community-oriented) to reflexive (self-actualization) volunteering \cite{Hustinx2003}, perhaps in response to the growing role of open-source participation in career development 
	\item Increasing code complexity requiring skills fewer developers possess and discouraging newcomers \cite{Steinmacher2015}
        \item Increasing formalization of software projects requiring significant effort on the part of developers to adhere to submission or foundation guidelines, similar to what has been observed with Wikipedia \cite{Jullien2015}
	\item A decreased quality of contributions and, therefore, a lower acceptance rate of contributions and an overload for reviewers and committers
	\item A saturation in quality and functionality for open source
\end{itemize}

The data are not sufficient for a robust prediction model on the growth of open source: On the one hand, we could not overcome concerns towards the measurement accuracy and precision of our measurement system Open Hub (Observations \ref{obs_new_projects_drop} and \ref{obs_erratic_contributor_rise}, unclear project duplication detection, unknown bot activity) and unclear representativeness of our data set. On the other hand, open source could have reached a local or global maximum, the growth could be best described as bell-shaped or does not follow any regular pattern.

\section{Conclusion}

In this study, we conducted a large-scale sample study on open-source projects and their cumulative growth. We analyzed the number of developers and their contributions with respect to lines of code, commits, and new projects to the open-source universe. We leveraged Open Hub as a measuring system to measure development activities of 172,833 open-source projects over the last 30 years concerning those four quantities. 

We could confirm an initial, transient exponential growth as claimed by Study~C \cite{Deshpande2008}. However, none of our accumulated measurements on lines of code, commits, contributors, or the number of projects remained exponential, quadratic, or linear in terms of growth as suggested by the prior studies. In fact, we already passed the peak in 2013 and observed a downwards trend for the measurements on commits and contributors in our data set.

Still, the greatest weakness of our study is the same as for the replicated studies: the underlying data source---in our case Open Hub. Although other studies \cite{Deshpande2008, Nagappan2013} assumed Open Hub data to be representative and the alternative data sets available have their own limitations, we are still stuck for an answer if the open-source project universe observed by Open Hub is a representative sample for open source or not. 

Thus, we encourage other researchers to replicate our study using our analysis pipeline and the complete data set\footnote{We published all Python analysis scripts and the full data set in raw and preprocessed state to the extent the data license allows under \url{https://github.com/michaeldorner/quo-vadis-open-source}} and pipeline as well as using utilizing a different data source. 


\section*{Acknowledgments}

We would like to thank Peter Degen-Portnoy for his support during the data collection, Synopsis for the data release, and Julian Frattini for his comments that greatly improved the manuscript. We would also like to thank Dirk Riehle for encouraging this research. This work was supported in part by the KKS Foundation through the SERT Research Profile project at Blekinge Institute of Technology.

\bibliographystyle{elsarticle-harv}
\bibliography{references.bib}

\end{document}